\documentclass[prb,twocolumn,superscriptaddress,showpacs,preprintnumbers]{revtex4}
\usepackage{graphicx,amssymb,color}
\usepackage{dcolumn}
\usepackage{bm}

\begin{document}
\draft \preprint{Lee}

\title {Electronic structure and Magnetic Properties of
Epitaxial FeRh(001) ultra-thin films on W(100)}

\author{J.-S. Lee}
\affiliation{National Synchrotron Light Source, Brookhaven National
Laboratory, Upton, New York 11973, USA}

\author{E. Vescovo}
\affiliation{National Synchrotron Light Source, Brookhaven National
Laboratory, Upton, New York 11973, USA}

\author{L. Plucinski}
\affiliation{Institute of Solid State Research IFF-9, Research
Centre J\"{u}lich, J\"{u}lich, Germany}

\author{C. M. Schneider}
\affiliation{Institute of Solid State Research IFF-9, Research
Centre J\"{u}lich, J\"{u}lich, Germany}

\author{C.-C. Kao} \thanks{{\it Current address}:  Stanford Synchrotron
Radiation Light source, SLAC National Accelerator Laboratory, Menlo
Park, CA 94025, USA} \affiliation{National Synchrotron Light Source,
Brookhaven National Laboratory, Upton, New York 11973, USA}

\date{\today}

\begin{abstract}
Epitaxial FeRh(100) films (CsCl structure, $\sim 10\ ML\ $ thick),
prepared {\it in-situ} on a W(100) single crystal substrate, have
been investigated via valence band and core level photoemission. The
presence of the temperature-induced, first-order, antiferromagnetic
to ferromagnetic (AF$\rightarrow$ FM) transition in these films has
been verified via linear dichroism in photoemission from the Fe 3$p$
levels. Core level spectra indicate a large moment on the Fe atom,
practically unchanged in the FM and AF phases. Judging from the
valence band spectra, the metamagnetic transition takes place
without substantial modification of the electronic structure. In the
FM phase, the spin-resolved spectra compare satisfactorily to the
calculated spin-polarized bulk band structure.
\end{abstract}

 \maketitle

\section{Introduction}

Stoichiometric 50:50 FeRh alloy has been studied extensively for
many years. Fallot\cite{Fallot38} reported on the discovery of an AF
$\rightarrow$ FM transition in this material more than half a
century ago. The unusual occurrence of a metamagnetic transition
attracted considerable attention and many valuable information have
been gathered over the years.

Equiatomic FeRh crystallizes in the CsCl structure. The magnetic
transition happens abruptly as a function of temperature (first
order transition, $T_{AF \rightarrow FM}$ $\sim$ 350K). The critical
temperature is strongly dependent on the stoichiometry (the
transition exists only within $\sim 5$\% deviation from the
equiatomic composition) as well as on the presence of transition
metals (TM) impurities. The transition is isostructural, but
accompanied by a rather large lattice constant
increase.\cite{Zsaldos67} The transition can also be reversed by
application of external pressure.\cite{Vinokurova76,Kulikov92} Other
physical quantities undergoing large variations at the transition
are thermal capacity, electrical resistivity and
entropy.\cite{Kouvel66}

Magnetically, the antiferromagnetic phase has been identified via
neutron scattering as of type II (AF-II), with nearest-neighbor Fe
sites ordered antiferromagnetically in the (001)-planes and
ferromagnetically in (111)-planes.\cite{Shirane64} Neutron data also
provide evidence of an enhanced Fe magnetic moment in this compound
($\sim 3\mu_B$) compared to bulk Fe metal ($\sim 2.2\mu_B$).
According to electronic structure calculations, enhanced Fe moments
are present in both the AF and FM phases. The Rh atoms are usually
assumed to be unpolarized in the antiferromagnetic phase and to
acquire $\sim 1\mu_B$ in the ferromagnetic phase.

Several models have been proposed to account for the properties of
the metamagnetic transition, but the debate is still active. Early
models, focusing on the conspicuous lattice expansion, attempted to
justify the transition in terms of an underlying lattice
instability. These theories encounter serious difficulties in
explaining the large entropy jump associated with the transition. On
the other side, the large variation of thermal and electrical
conductivities motivated the search for an electronic origin for the
change in entropy. This search generally led to theoretical models
attributing a major role to modifications of the electronic
structures (i.e. change in density of states, DOS) in the AF and FM
phases.

Recently, interest in the FeRh system has revived. Its peculiar
thermo-magnetic properties promise the realization of efficient
thermally assisted magnetic recording
media.\cite{Thiele03,Thiele04,Guslienko04} With this new interest,
the metamagnetic transition has been re-scrutinized using modern
experimental techniques. Notably, time-resolved investigations have
attempted to elucidate the role played by the spin-lattice dynamics
in the transition.\cite{Ju04,Thiele2004}

In spite of the richness of information on the properties of FeRh,
direct spectroscopic information on this system is conspicuously
scarce, especially for ultra-thin films. X-ray magnetic dichroism
studies\cite{Stamm08,Chaboy1999} from polycrystalline samples
focused on the magnetic properties without examining effects in the
electronic structure, aside from the XMCD signal in the FM phase.

In this work, we report on a comprehensive photoemission study of
FeRh ultrathin films with close-to-equiatomic compositions.
Epitaxial FeRh(001) films are obtained by room-temperature
co-deposition on a W(100) single crystal substrata, followed by mild
annealing. The atomic reordering process, induced by post-annealing,
is accompanied by substantial modifications of the electronic
structure, apparent in core levels as well as in the valence band
spectra. The appearance of a cubic LEED pattern at this stage
signals the formation of a well-defined, single crystalline alloy
films. The temperature dependence of the magnetic properties are
monitored using magnetic linear dichroism (MLD) in the Fe 3$p$ core
levels. They reveal that the AF $\rightarrow$ FM metamagnetic
transition, characteristic of bulk equiatomic alloy, is present also
for these ultra-thin epitaxial films and in particular at their
surfaces. In the ferromagnetic phase, spin-resolved valence band
spectra display well polarized features which are consistent with
\emph{first-principle} calculations of the spin-polarized band
structure of bulk ferromagnetic FeRh. Judging from the photoemission
spectra, the metamagnetic transition is accomplished without any
major change in the electronic structure. In particular, large
modifications of the $d$-derived DOS can be excluded during the
metamagnetic transition.

\section{Experimental Details}

All experiments were performed at beamline U5UA at the National
Synchrotron Light Source (NSLS). The beamline, equipped with a
planar undulator and a high-resolution spherical-grating
monochromator, is dedicated to spin-polarized angular-resolved
photoemission (SP-ARPES) studies.\cite{Vescovo99}

The incident light was linearly polarized in the horizontal plane. A
commercial hemispherical electron energy analyzer, coupled to a
mini-Mott polarimeter for spin analysis, was employed for these
experiments. The incoming photons, incident at 45$^\circ$ from the
sample normal, and the emitted (detected) electrons were in the
horizontal plane. The emission angle was along the sample surface
normal. The typical overall instrumental resolution was
approximately 80 meV for spin integrated and 120 meV for
spin-polarized photoemission spectra. The Sherman function of the
spin-detector was estimated at 0.2. The FeRh films were magnetized
along the \emph{in-plane} [010] (vertical) direction via a pulsed
current passed through a magnetizing coil placed near the sample
surface. All magnetic measurements were performed in magnetic
remanence.

\begin{figure}[t]
\begin{center}
\includegraphics[width=0.47\textwidth] {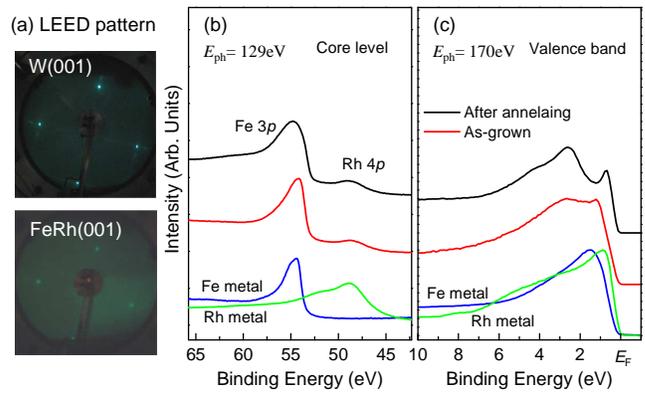}
\caption{(color online) (a) LEED patterns from W(001) (top, $E_{\rm
ph}$ = 54 eV) and from an epitaxial film (10 ML) of FeRh(001)
(bottom; $E_{\rm ph}$ = 57.2 eV). (b) Photoemission spectra from
shallow core levels (Fe 3$p$ and Rh 4$p$) from equiatomic FeRh
films, before and after annealing. The spectra from films of Fe and
Rh metals are also shown for comparison. (c) Valence band spectra
corresponding to the spectra shown in (b).} \label{Alloy-formation}
\end{center}
\end{figure}

FeRh(001) films ($\sim$ 10ML) were grown on a W(001) single crystal
at RT by co-evaporation of Fe and Rh from two electron beam sources.
The evaporation rate was $\sim$ 5 $\rm \AA$/min. The base pressure
in the chamber was 5$\times$10$^{-11}$ Torr and rose to about
1$\times$10$^{-10}$ Torr during depositions. After RT deposition,
the LEED was barely visible, consisting of large, diffused spots on
a high background. Annealing to 600$^\circ \rm{C}$ for approximately
2min restored a sharp cubic LEED pattern (see Fig.\
\ref{Alloy-formation}(a)). The pattern is in registry with the
underlying W(001) substrate, indicating the formation of ordered
films, exposing the (001) surface. The annealing procedure has been
limited because longer/higher annealing leads to substantial three
dimensional island growth with partial uncovering of the substrate,
as can be assessed by monitoring the W 4$f$ peaks.

\section{Photoemission Results and Discussion}

The formation of the ordered alloy is accompanied by considerable
modifications in the electronic structure, easily detected in the
photoemission spectra from core levels and the valence band (see
Fig.\ \ref{Alloy-formation}(b) and \ref{Alloy-formation}(c)).

At the 50:50 concentration the Fe 3$p$ signal is more intense than
Rh 4$p$. This is due to a combination of favorable Fe photoemission
cross sections at 130 eV ($\sigma$(Fe$_{3p}$)/$\sigma$(Rh$_{4p}$)
$\sim$2) and to a Rh signal spread over a wider energy range as a
consequence of the larger spin-orbit coupling in the heavier
element. In the {\it as-grown} films, the line-shape of the Fe 3$p$
is similar to metallic Fe. During the annealing process, the 3$p$
spectrum broadens and shifts toward higher binding energy. These
features represent the substitution of Fe nearest-neighbors with Rh
atoms during the formation of the CsCl cubic structure of FeRh. The
annealing process does not change appreciably the Rh-to-Fe core
level intensity ratio suggesting a negligible tendency to surface
segregation, either of Fe or Rh. A more quantitative assessment
cannot be safely obtained from our data due to the change in line
shape of the Fe $3p$ and Rh $4p$ spectra in the alloy and in the
pure metals and to their partial overlap.

The sharpening of all spectral features in the valence band spectrum
upon annealing (see Fig.\ \ref{Alloy-formation}(c)) can also be
partially explained by the onset of long-range crystallographic
order, with the consequent decrease of inelastic scattering events
for the out-going electrons. In our case, however, a more
substantial charge redistribution is taking place. Consistently with
the formation of the FeRh alloy at this stage, the form of the
valence band is modified by the annealing process. A new peak
appears close to the Fermi level and higher spectral intensity
builds up at $\sim$ 5 eV binding energy. At the relatively high
photon energy used in these experiments ($E_{\rm ph}$ = 170 eV),
photoemission spectra represent, in first approximation, densities
of states, mostly biased towards $d$ states. The accumulation of
intensity at high binding energy can be attributed to 3$d$-4$d$
majority spin hybridization during alloy formation. Indeed the
majority $d$-bands largely overlap in Fe and Rh metals. By the same
mechanism, the occupied Fe minority spin states close to the Fermi
level are pushed upward to become unoccupied, anti-bonding states.
The result is an enhancement of the Fe magnetic moment, but also an
increased separation between the Fe and Rh minority spin states.
Largely unhybridized, Rh minority spin states just below the Fermi
level therefore account for the sharp peak appearing in the spectra
at approximately 0.5 eV binding energy.

\begin{figure}[b]
\begin{center}
\includegraphics[width=0.47\textwidth] {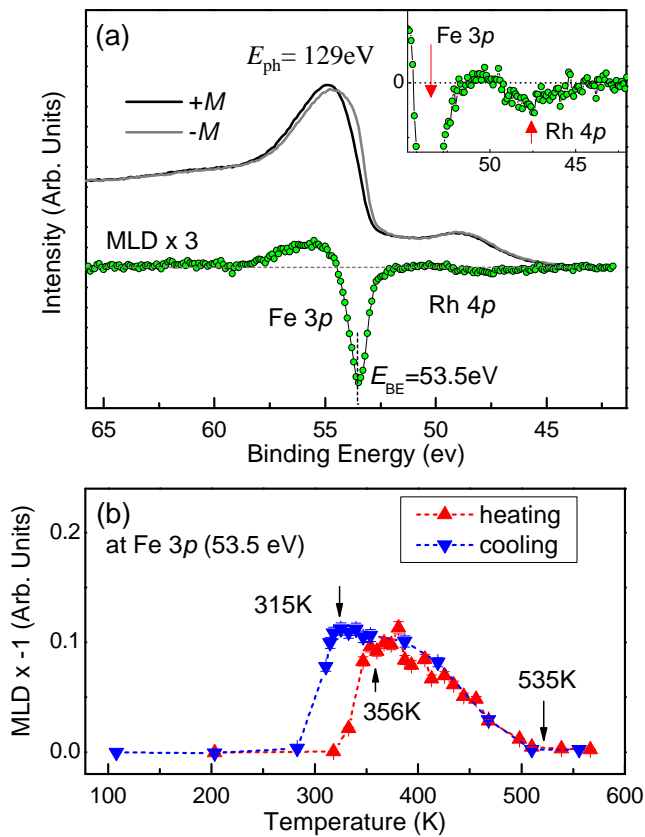}
\caption{(color online) (a) linear magnetic dichroism in
photoemission from Fe 3$p$ and Rh 4$p$ core levels of an FeRh film
in the ferromagnetic phase. Inset shows a blow up of the Rh $4p$
region. (b) temperature dependence of the Fe 3$p$ dichroism signal.}
\label{Magnetism}
\end{center}
\end{figure}

The magnetic behavior of the FeRh films can be monitored using
magnetic linear dichroism in core-level photoemission.\cite{Roth93}
These spectra are shown in Fig.\ \ref{Magnetism}(a) for FeRh film in
the ferromagnetic phase. The two continuous lines (black and gray)
correspond to the photoemission spectra measured with up and down
magnetization directions. The difference curve, the dichroism
signal, is also shown. Clearly the Fe dichroism signal is much
larger than the Rh one (see Fig.\ \ref{Magnetism}(a) inset). This is
consistent with the large difference in local magnetic moments on Fe
($\sim$3$\mu_B$) and Rh ($\sim$1$\mu_B$) sites. The similar shape of
the difference curves of the two dichroisms (within their binding
energy ranges) indicates the parallel alignment of the Fe and Rh
moments.

The temperature dependence of the Fe 3$p$ dichroism is shown in
Fig.\ \ref{Magnetism}(b) (bottom) in the range 100$\sim$600K. We
note that a pulsed magnetic field before each series of measurements
as a function of temperature was applied. It is clear that the
$AF\rightarrow FM$ transition is present in the ultrathin films
prepared in this study. The critical temperature of the metamagnetic
transition, separating the antiferromagnetic state (low temperature,
no dichroism) from the ferromagnetic state (high temperature, finite
dichroism), is $\sim$ 50K above room temperature. The main features
of the transitions -- e.g. the critical temperature and range of
hysteretic behavior\cite{hysteresis} -- are remarkably close to the
values found in bulk FeRh alloys. Only the steepness of the
transition increases in the single-crystalline films; possibly an
indication of the intrinsic first order nature of the transition.

It is relevant to recall that photoemission at the excitation
energies used in our experiments probes at most the first few
monolayers of material. These results therefore demonstrate that the
metamagnetic transition is present also at the FeRh surface.
Furthermore, considering the lattice strain usually associated with
epitaxial thin films -- particularly at their surfaces and
interfaces -- these observations suggests that a lattice instability
is a consequence rather than the driving force of the transition.

\begin{figure}[t]
\begin{center}
\includegraphics[width=0.45\textwidth] {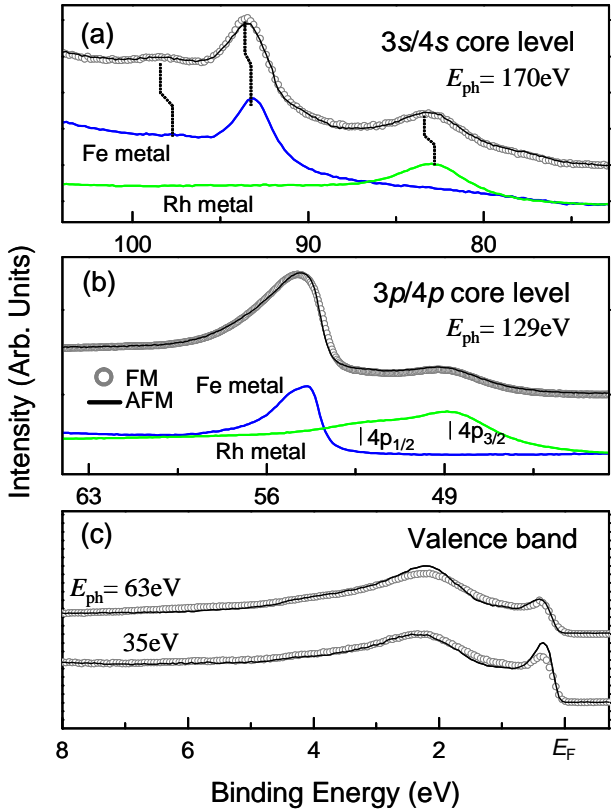}
\caption{(color online) (a) Fe 3$s$ and Rh 4$s$ core level spectra
from Fe and Rh metals and from equiatomic FeRh alloy in the AF
(solid line) and FM (open circles) phases; (b) similar to (a) but
for Fe 3$p$ and Rh 4$p$ levels; (c) normal emission valence band
spectra taken close to the high-symmetry $\Gamma$ and X point.}
\label{Transition}
\end{center}
\end{figure}

Above the metamagnetic transition, further increasing the
temperature, the magnetic signal diminishes. The Curie temperature,
above which the system is paramagnetic, is at approximately 530K. It
is interesting to note that at $\sim$ 430K the slope of the Curie
curve appears to change slightly, accelerating the decay of the
magnetization. This behavior, if further confirmed via a more direct
probe of the magnetization, could indicate the disappearance of the
local Rh moment via a high- to low-spin state transition. The
existence of these two states for the Rh atoms has been suggested in
a recent model proposed to explain the FeRh thermomagnetic
behavior.\cite{Gruner03}

\begin{figure}[t]
\begin{center}
\includegraphics[width=0.44\textwidth] {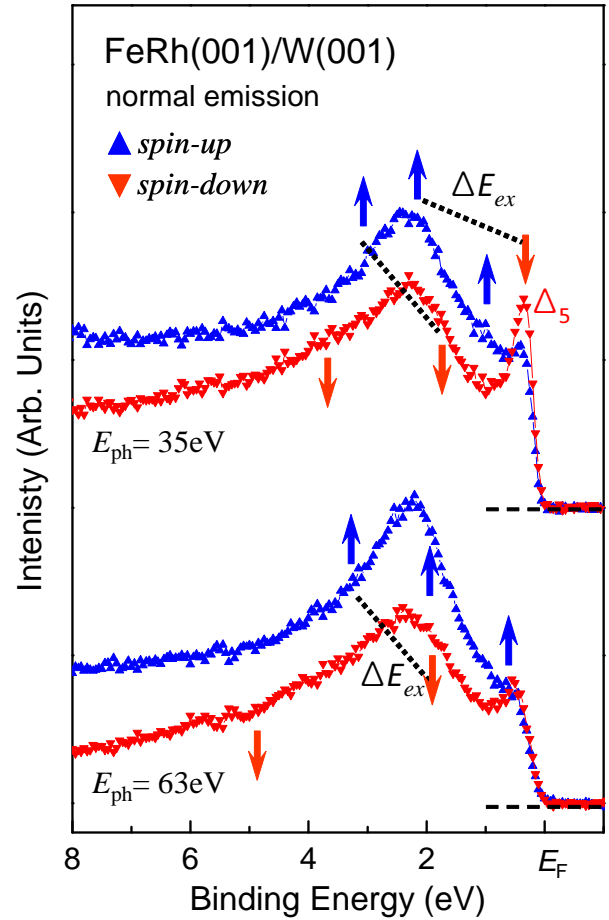}
\caption{(color online) Valence-band spin-resolved spectra in the
ferromagnetic phase at the high-symmetry $\Gamma$ and X points that
are marked with arrows (up/down: majority/minority). The exchange
splitting ($\Delta E_{ex}$) is estimated to be around 1.4 eV. See
text for details. The clearest feature is perhaps the $\Delta_5$
sharp minority state close to the Fermi level ($\sim$ 0.2 eV binding
energy) in the spectra.} \label{Spin}
\end{center}
\end{figure}

The effect of the AF$\rightarrow$FM transition on the electronic
structure is shown in Fig.\ \ref{Transition}. The photoemission
spectra from films in the antiferromagnetic state (continuous lines,
$\sim$100K) and in the ferromagnetic state (open circles,
$\sim$380K) are superimposed to facilitate the comparison. The
magnetic transition has remarkably little effect on all
photoemission spectra; core levels ($s$-levels, top; $p$-levels,
middle) as well as valence band (bottom). The small differences in
the spectra are mostly attributable to thermal broadening.

In Fig.\ \ref{Transition}(a), the 3$s$ and 4$s$ spectra of Fe and Rh
metals are compared to the FeRh one. The overall shift of these
levels ($\sim$0.5 eV) reflects an increased local electrostatic
potential in the alloy. Furthermore, the distance between the main
peak and the higher energy satellite of Fe 3$s$ increases from 4.5
eV in the metal to 5.1 eV in the alloy. Due to the absence of
orbital contributions in the $s$-levels, the distance between the
main peak and the higher energy satellite is proportional to the
local spin moment on the Fe site.\cite{Kachel93} This observation
therefore implies a larger Fe magnetic moment in the alloy.

The shape of the Fe 3$s$ core level remains essentially unchanged
during the metamagnetic transition. A larger magnetic moment is
therefore present in both phases of the alloy. These observations
agree with calculations of the electronic structure which give a Fe
magnetic moment of approximately 3 $\mu_B$ in FeRh to be compared
with 2.2 $\mu_B$ of Fe metal. Unfortunately, the signal from the Rh
4$s$ level is too weak to permit an unambiguous assessment on the
presence of a high energy satellite, as would be expected during the
formation of a Rh magnetic moment in the FM state. Similarly, as in
the case of the $s$-levels, the $p$-levels (Fig.\
\ref{Transition}(b)) are also mostly unaffected by the metamagnetic
transition.

\begin{figure*}
\begin{center}
\includegraphics[width=0.87\textwidth] {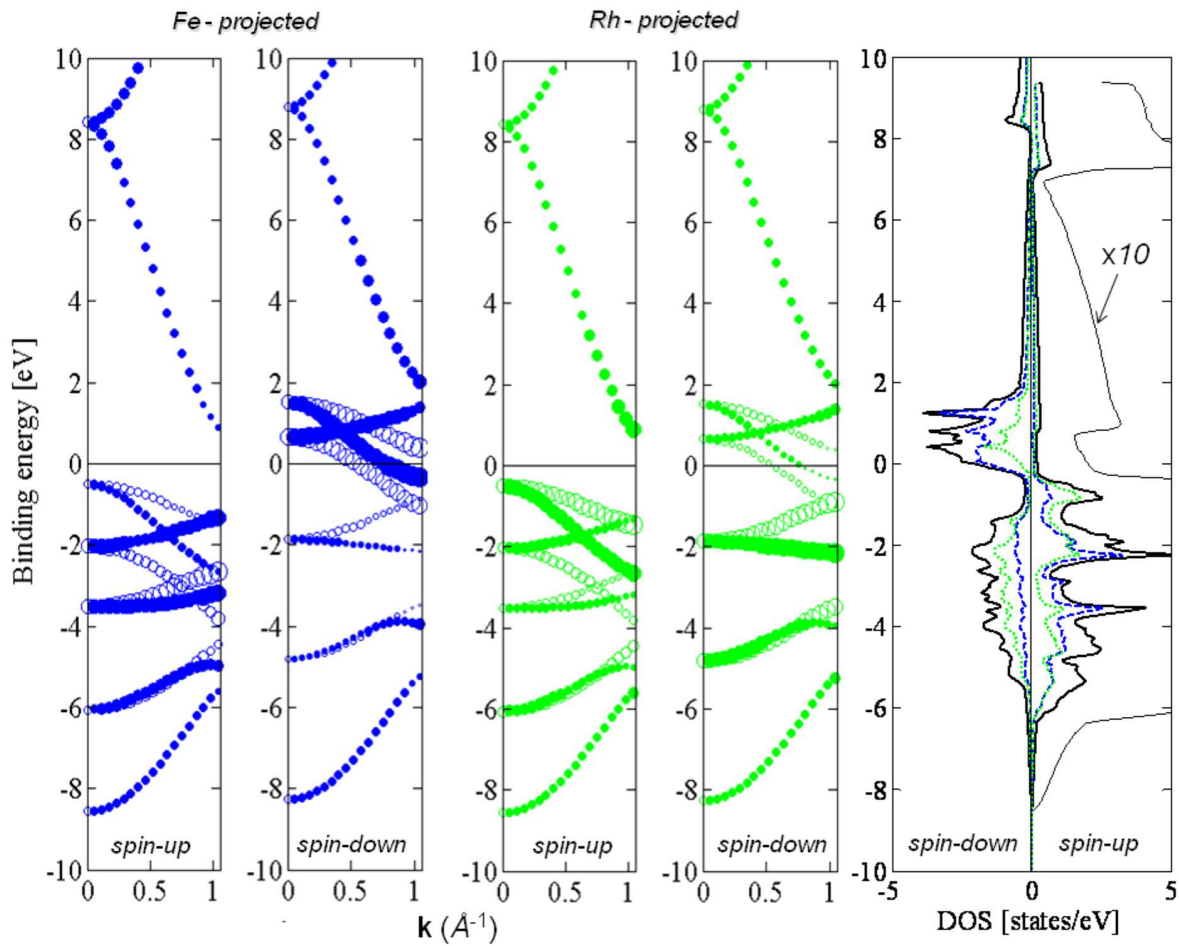}
\caption{(color online) Electronic bands and DOS of ferromagnetic
FeRh. Only bands along the $\Delta$ high-symmetry line ($\Gamma$-X
line) of the $fcc$ BZ are shown, full circles indicate bands of
$\Delta_{1}$ or $\Delta_{5}$ symmetries. See text for details.}
\label{Band-structure}
\end{center}
\end{figure*}

The angular resolved photoemission spectra from the valence band are
shown in Fig.\ \ref{Transition}(c). The spectra are measured in
normal emission (i.e. ${\bf k}_{\parallel}$=0) from the (100)
surface, and probe states along the bulk $\Delta$ line. The two
photon energies shown sample widely spaced regions of the Brillouin
zone (BZ). Assuming a free electron final state, 63 and 35 eV photon
energies correspond to states close to the $\Gamma$-point and
X-point, respectively. Similarly to the core levels, the valence
band remains largely unchanged during the metamagnetic transition.
Again the major effect is thermal broadening, easily visible at the
Fermi level. This behavior is not limited to the two photon energies
shown; we have explored the photon energy range between 20 to 80 eV
and found essentially the same result.

The spin-polarized electronic structure in the FM phase can be
examined with spin-resolved photoemission. The spin-resolved valence
band spectra, measured under the same conditions as Fig.\
\ref{Transition}(c) are shown in Fig.\ \ref{Spin}. These well
polarized spectra are quite complex, displaying several overlapping
features in both minority and majority spin channels.

As a first step in interpreting these spectra, it is useful a
comparison with band structure calculations. In magnetic 3$d$
metals, however, electronic correlation have been shown to affect
the photoemission spectra\cite{correlation1,correlation2} and a more
refined analysis would need to take them into proper account. The
band structure caculation were performed using the WIEN2k
package,\cite{WIEN2k} with default settings and a lattice constant a
= 2.985 \AA.\cite{Koenig81}

The band structure along the high symmetry $\Delta$-line is
displayed in Fig.\ \ref{Band-structure}. The general form of the
bands resembles a typical bcc elemental ferromagnet (e.g. Fe metal),
doubly folded because of the two elements in FeRh.\cite{note2} This
folding is indeed the reason why the photoemission spectra contains
so many features.

Figure\ \ref{Band-structure} also shows spin-resolved density of
states (DOS) which was calculated with a 21$\times$21$\times$21 mesh
in the BZ, with effective 286 \emph{k}-points in the irreducible BZ,
and plotted with 40 meV Gaussian broadening. The general shape of
the DOS and its magnitude compare favorably with previous
calculations,\cite{Koenig81,Moruzzi92} with small alterations most
likely related to the different computation methods used.

In the experimental geometry used, only initial states of
$\Delta_{1}$ or $\Delta_{5}$ symmetry are
dipole-allowed.\cite{Turner83} Their positions at the two
high-symmetry points, $\Gamma$ and X, are marked with arrows
(up/down: majority/minority) in Fig.\ \ref{Spin}. Although the
presence of several broad and overlapping features do not allow for
an unambiguous assignment of all features, there is generally a good
correspondence between the experimental features and calculated
bands. The clearest feature is perhaps the $\Delta_5$ sharp minority
state close to the Fermi level ($\sim$0.2 eV binding energy) in the
spectra at 35 eV, which is not present at 63 eV. Along the
$\Delta$-line the $\Delta_5$ band disperses downward, being
unoccupied at $\Gamma$ and occupied at X (see Fig.\
\ref{Band-structure}). The majority counterpart of this state is
located at a binding energy of $\sim$ 2.5 eV, giving a rather large
exchange splitting ($\Delta E_{ex}$ = 2.3 eV, see Fig.\ \ref{Spin}).

The 3$d$-4$d$ hybridization in this system results in an unusually
variable exchange. Besides states with high exchange splitting,
other $d$ states have a much lower exchange. An example of this are
the minority and majority states located at 1.8 and 3.2 eV,
respectively; giving an exchange splitting of only 1.4 eV (see Fig.\
\ref{Spin}). It would be tempting to attribute states of high and
low exchange to Fe and Rh states, respectively. This picture however
is oversimplified as can be concluded from the site-projected
calculations shown in Fig.\ \ref{Band-structure} where bands are
highlighted according to the weight on Fe-sites or on Rh-sites. The
main features emerging from Fig.\ \ref{Band-structure} are that
majority bands are roughly equi-distributed on the Fe and Rh sites
while the minority bands are split into two groups: Rh-like in the
lower portion and Fe-like close to the Fermi level. This behavior
stems from the different degree of hybridization of majority and
minority states mentioned above. The main factor determining the
charge distribution in the alloy is the energy location of the bands
previous to the alloy formation rather than the spatial
localization. The majority bands of Fe and Rh, largely degenerate,
strongly hybridize forming delocalized states equally distributed on
the Fe and Rh sites. The minority bands, hybridizing much less,
remain localized, forming bonding states on the Rh sites and
anti-bonding states on the Fe sites.\cite{note} It is therefore
meaningless to attribute exchange pairs to either Fe or Rh.

As mentioned in the introduction, there is a vast literature on the
origin of the metamagnetic transition in FeRh. It is therefore worth
to discuss our results in the context of previously published work.

A primary feature of our experimental work is the investigation of
epitaxial films. The metamagnetic transition is accompanied by a
relatively large (in the range of $\sim$ 0.5--1\%)\cite{Kouvel} and
isotropic lattice expansion. Moreover, the transition temperature
can be varied and the transition itself can even be suppressed by
external pressure applied to FeRh samples. These observations have
lead some researchers to link the transition to a lattice
instability, somewhat along the lines of the Invar phenomena in
FeNi.\cite{Pepperhoff} It seems therefore appropriate to attempt
FeRh experiments on single crystalline samples (epitaxial films)
consisting of only a few monolayers ($\sim$ 10 ML in the present
films). The FeRh bulk lattice constant is 2.99 $\rm \AA$ while the W
lattice parameter is 3.15 $\rm \AA$. The anticipated consequence of
this 5\% mismatch is the substantial lattice strain if pseudomorphic
growth is assumed. The fact that the magnetic properties of these
ultrathin films are essentially the same as those of much thicker
polycrystalline films, or even bulk FeRh, is a strong indication
that the lattice change may not be a decisive factor of the
metamagnetic transition, but rather a consequence of it.

The considerable change in the specific heat and resistivity at the
metamagnetic transition had been noticed early on in the study of
FeRh.\cite{Ivarsson71} These measurements are challenging and often
have to rely on comparison of samples with slightly different
stoichiometry. Nevertheless, these experimental results suggest
profound changes in the electronic structure. Redistribution of
charge could account for large changes of the electrical and thermal
properties of the material undergoing the metamagnetic
transition.\cite{Baranov95,Annaorazov95} Indeed, early electronic
structure calculations displayed a large change of the density of
states across the transition.\cite{Koenig81} However, core level,
high-energy valence band (XPS), and low-energy valence band (ARPES)
photoemission spectra do not show any significant differences across
the transition. Although the relatively high temperature needed to
reach the FM phase limits the resolution to about 200 meV, at least
within this limit, large charge rearrangements can be excluded. The
direct sampling of the electronic structure using photoemission
therefore substantiates and extends previous findings obtained with
ellipsometry measurements\cite{Chen88} and with positron
annihilation techniques\cite{Adam72} on bulk samples. They also
agree with modern first principle calculations of the electronic
structure which find similar charge transfer from Rh to Fe in both
magnetic phases as well as similar values for both majority and
minority spin density of states at the Fermi energy.\cite{Moruzzi92}

\section{Summary}

In conclusion, equiatomic FeRh(100) films (CsCl structure, $\sim$ 10
ML) have been epitaxially grown on W(100) by room temperature
co-deposition of Fe and Rh, followed by moderate annealing. The
formation of the ordered alloy is accompanied by profound
modifications of the electronic structure, displayed in valence band
and shallow core level photoemission spectra. The epitaxial
ultra-thin films and particularly their surfaces display magnetic
properties similar to bulk FeRh. The temperature-induced $AF
\rightarrow FM$ transition is clearly detected via magnetic linear
dichroism in photoemission. Surprisingly, the metamagnetic
transition is accomplished without major changes in the electronic
structure: valence band and core levels remain practically
unaffected crossing the critical temperature. In the FM phase, the
valence band spectra display well polarized features, in good
agreement with spin-polarized band structure calculations.

\section*{ACKNOWLEDGEMENTS}

The NSLS, Brookhaven National Laboratory, is supported by the U.S.
DOE, Office of Science, Office of Basic Energy Sciences, under
Contract DE-AC02-98CH10886.




\begin{thebibliography}{}
\bibitem{note2} The FeRh CsCl-structure can be thought as a
simple bcc structure with every central atom substituted with the
other element. In first approximation, the resulting band structure
is therefore the one of a bcc crystal with an average electrostatic
potential between Fe and Rh. The doubling of the lattice
periodicity, due to the different elements, additionally results in
folding the bands in reciprocal space.
\bibitem{note} The behavior shown in Fig. 5 along the
$\Delta$-line is confirmed along all high-symmetry lines of the
Brillouin zone.
\end{thebibliography}
\end{document}